\documentclass[preprint2]{aastex}
\usepackage{apjfonts}
\usepackage{psfig}

\newlength{\abstrwidth}
\def\baselinestretch{0.92}
\textheight=\baselinestretch\textheight
\ifdim\textheight>24.6cm\textheight=24.4cm\fi

\slugcomment{Revised 14 August 2000}

\lefthead{Boyd et al.}
\righthead{The low/hard state of LMC X--3}

\def\source{LMC~X--3}
\def\approxlt{\mathrel{\hbox{\rlap{\lower.55ex \hbox {$\sim$}}
        \kern-.3em \raise.4ex \hbox{$<$}}}}
\def\approxgt{\mathrel{\hbox{\rlap{\lower.55ex \hbox {$\sim$}}
        \kern-.3em \raise.4ex \hbox{$>$}}}}

\def\kms{$\,$km$\,$s$^{-1}$}

\begin{document}

\title{Canonical Timing and Spectral Behavior of LMC X-3 in the Low/Hard State}

\author{Patricia T. Boyd \altaffilmark{1,2},
Alan P. Smale \altaffilmark{1}}

\affil{Laboratory for High Energy Astrophysics,
Code 662, NASA/Goddard Space Flight Center, Greenbelt, MD 20771}

\altaffiltext{1}{Also Universities Space Research Association.} 
\altaffiltext{2}{Also Joint Center for Astrophysics, University of
Maryland, Baltimore County}

\author{Jeroen Homan \altaffilmark{3},
Peter G. Jonker \altaffilmark{3},
Michiel van der Klis \altaffilmark{3}}

\affil{Astronomical Institute `Anton Pannekoek', University of
Amsterdam}

\altaffiltext{3}{Also Center for High-Energy Astrophysics, 
Kruislaan 403, 1098 SJ, Amsterdam}

\author{Erik Kuulkers \altaffilmark{4}}

\affil{
Space Research Organization Netherlands, Utrecht}

\altaffiltext{4}{Also Astronomical Institute, Utrecht University}

\begin{abstract}

We present results from three observations with the Rossi X-ray Timing
Explorer (RXTE) of LMC X-3, obtained while the source was in an
extended 'low/hard' state. The data reveal a hard X-ray spectrum which
is well fit by a pure power law with photon index
$\Gamma$=1.69$\pm$0.02, with a source luminosity at 50~kpc of
5--16$\times$10$^{36}$ erg~s$^{-1}$ (2--10 keV). Strong broad-band
(0.01-100~Hz) time variability is also observed, with fractional rms
amplitude  40$\pm$4\%, plus a quasi-periodic oscillation (QPO)
peak at 0.46$\pm$0.02 Hz with rms amplitude $\sim$14\%.
This is the first reported observation in which the full canonical
low/hard state behavior (pure hard power law spectrum combined with
strong broad-band noise and QPO) for \source\ is seen.  We reanalyze
several archival RXTE observations of LMC X-3 and derive consistent
spectral and timing parameters, and determine the overall luminosity
variation between high/soft and low/hard states. The timing and
spectral properties of LMC X-3 during the recurrent low/hard states
are quantitatively similar to that typically seen in the Galactic
black hole candidates.

\end{abstract}


\section{Introduction}

Galactic X-ray binaries harboring a black hole candidate (BHC), such
as Cygnus X--1 and GX~339--4, exhibit a number of distinguishable
states characterized in terms of total luminosity, energy
spectral parameters, and time variability (see e.g.\ reviews by van
der Klis 1995, Tanaka \& Lewin 1995, Nowak 1995, and references
therein).  During the high/soft state, the 2--10 keV spectrum can be
modeled with a significant thermal component, and the rms time
variability of the power spectral density is only a few percent. The
more typical (for Galactic systems) low/hard state is well described
by a non-thermal spectrum, represented by a power law with photon
index $\sim$1.7, and significant time variability
with rms amplitudes of 30--50\%. QPOs seen in this
state have typical frequencies of 0.2--3~Hz (Wijnands \& van der Klis
1999).
 
\source, a bright (up to 3$\times$10$^{38}$ erg~s$^{-1}$) BHC in the
Large Magellanic Cloud, is highly variable on timescales from days to
years. It is typically observed in the high/soft state, with an X-ray
spectrum qualitatively similar to that of other BHCs in the soft
state: an "ultrasoft" component and a hard ($>$10keV) tail (White \&
Marshall 1984).  The ultrasoft component is well represented by an
optically-thick accretion disk model (Shakura \& Sunyaev 1973) with
$kT$ = $\sim$ 1.1 keV and a variable mass accretion rate.  The B3~V (V
$\sim$16.7-17.5) optical counterpart shows a large velocity range with
semi-amplitude K=235 \kms\ 
through its 1.7-day orbital period. The lack
of eclipses implies an orbital inclination of $<$70$^\circ$ and a
compact object mass of $\sim$7M$_{\sun}$ (Cowley et al.\ 1983,
Paczynski 1983, Ebisawa et al.\ 1993, but see also Mazeh et al.\
1986).
Cowley et al. (1991) presented evidence for a long-term periodicity of
$\sim$198 (or perhaps $\sim$99) days based on HEAO I and Ginga
observations. 
This variability was attributed to the precession of a
bright, tilted, and warped accretion disk.
Later ASM observations reveal a much more complex, and
less periodic, behavior (Nowak et al. 2000, Paul et al. 2000, Boyd
2000.)  

\source\ has been the subject of two monitoring campaigns
with RXTE, spanning 1996 February 2 through 1999 August 31.
The first
consisted of short $\sim$1~ks pointings, separated by several days;
the second used longer ($\sim$8--10~ks) pointings separated by 3--4
weeks. Based on these data, Wilms et al. (2000) report the discovery of
transitions from the high/soft to the low/hard state; during the
observation with lowest countrate, the disk
component vanishes and the spectrum can be fit by a pure power
law. 
This implies that a state transition, rather
than the periodically changing absorption column arising 
from a tilted, precessing disk, may be responsible for
the low-flux episodes of \source. Wilms et al. (2000)
suggest a model in which a wind-driven limit cycle gives rise to the
long term variability.

The RXTE ASM light curve of \source\ showed that a
possible low/hard state that began around 2000 April 10 was lasting
longer than typical.  We therefore arranged Target of Opportunity RXTE
observations to search for the
characteristic time variability of BHCs in the low/hard
state 
(Boyd \& Smale 2000, Homan et al. 2000).
We present the first
analysis in which the full canonical low/hard state behavior for
\source\ is seen. This is the first time such behavior has been
observed for a BHC outside our Galaxy, as well as being the first
detection of QPO in \source. 
Our analysis of archival data shows that such low/hard states are
recurrent in \source.

\section{Observations and Analysis}

Observations were performed with the RXTE satellite
(Bradt, Rothschild, \& Swank, 1993) on 2000 May 3, 10 and 13 for a
total onsource good time of 10.4 ksec (see Table 1).  The PCA
instrument on RXTE consists of five Xe proportional counter units
(PCUs), with a combined effective area of about 6500 cm$^2$ (Jahoda et
al 1996). In each of the three observations, 4 PCUs were collecting
data. We present results using the Standard 2,
E\_500us\_64M\_0\_1s, and Good Xenon data modes, with effective time
resolutions of 16 s, 500$\mu$s and $<1\mu$s respectively.

The spectral data were analyzed using 
FTOOLS 5.0.  Background subtraction was
performed using the faint source model (``L7-240'',
v19991214). We analyzed data from the top layer only,
to increase signal to noise.  Response matrices were generated using
PCARSP 2.43 with the latest energy-to-channel relationship (e04v01).
Spectral
fitting was performed using XSPEC 11.0.  We ignored
data below 2.5 keV and above 25 keV.

\begin{deluxetable}{lllllll}
\tablewidth{0pc}
\tablecaption{LMC X-3 RXTE Low/Hard State Observations}
\tablehead{
\colhead{Date} &
\colhead{Exp. (s)} &
\colhead{count rate/PCU} &
\colhead{$\Gamma$} &
\colhead{$A_{pl}$ } &
\colhead{$\chi^2$/dof} &
\colhead{$L_x$ (ergs/s)}
}
\startdata
1996 Jun 29  &   1056 & 2.25 &  1.83$^{+0.08}_{-0.08}$ \ \ \ \ \ \ & 6.30$^{+0.97}_{-0
.84}$ $\times$10$^{-3}$ &   26.8/32 \ \ \ \ \ &  6.3$^{+0.9}_{-0.8}$ $\times$10$
^{36}$  \nl
1996 Jul 04  &    976 & 0.76 &  1.57$^{+0.22}_{-0.25}$  &  1.29$^{+0.63}_{-0.49}$  &  
 23.8/32 &  1.9$^{+0.9}_{-0.7}$  \nl
1996 Jul 10  &    896 &  0.76 & 1.80$^{+0.16}_{-0.16}$  &  2.52$^{+0.80}_{-0.64}$  &  
 30.6/32 &  2.6$^{+0.8}_{-0.7}$  \nl
             &        &                     &                    &           &  
                  \nl
1998 May 29  &   6400 & 2.15 &  1.82$^{+0.03}_{-0.03}$  &  6.06$^{+0.30}_{-0.29}$  &  
 21.7/32 &  6.1$^{+0.3}_{-0.3}$  \nl
1998 May 29  &   3568 & 2.20 &  1.72$^{+0.05}_{-0.05}$  &  4.86$^{+0.42}_{-0.40}$  &  
 26.0/32 &  5.8$^{+0.5}_{-0.5}$  \nl
             &        &                     &                    &           &  
                  \nl
2000 May 05  &   1712 & 2.56 &  1.60$^{+0.04}_{-0.04}$  &  6.01$^{+0.45}_{-0.41}$  &  
 36.7/52 &  8.6$^{+0.6}_{-0.6}$  \nl
2000 May 10  &   2024 & 1.5 &  1.60$^{+0.05}_{-0.04}$  &  3.30$^{+0.34}_{-0.26}$  &  
 32.6/52 &  4.7$^{+0.5}_{-0.4}$  \nl
2000 May 13  &   6656 & 4.86 &  1.69$^{+0.01}_{-0.01}$  &  12.7$^{+0.26}_{-0.26}$  &  
 65.2/55 &  15.7$^{+0.3}_{-0.3}$  \nl
\enddata
\end{deluxetable}

We created power spectra using the high time
resolution data modes in three energy bands: 3--20 keV, 3--10 keV and
10--20 keV.  No background and deadtime corrections were applied.
The power spectra were
generated from 256s data segments, using a Nyquist frequency of 
1024 Hz. The individual power spectra were averaged and rms
normalized (Belloni \& Hasinger 1990, Miyamoto et al. 1991). The
Poisson level, determined by taking the unweighted average
of all powers between 500 and 1000 Hz, was subtracted from the power
spectrum. The resulting power spectrum was rebinned logarithmically
(to 60 frequency bins per decade) and fitted in the 1/256--256 Hz
range.  Errors on the fit
parameters were determined using $\Delta\chi^2=1$ (1$\sigma$ for a
single parameter of interest).  For observations with low count rates
($<$5 counts/s/PCU) the uncertainty in the background estimation
introduces an additional fractional error of 5--10\% in the rms
amplitudes. The errors quoted here are only the statistical ones.

\section{Spectral Results}

The long-term light curve of \source\ as measured by the All-Sky
Monitor (ASM) aboard RXTE is shown in Figure 1. 
Locations of the low/hard and high/soft state PCA observations
discussed below are indicated with vertical dashes.

\begin{figure}[t]
\figurenum{1}
\begin{center}
{\psfig{file=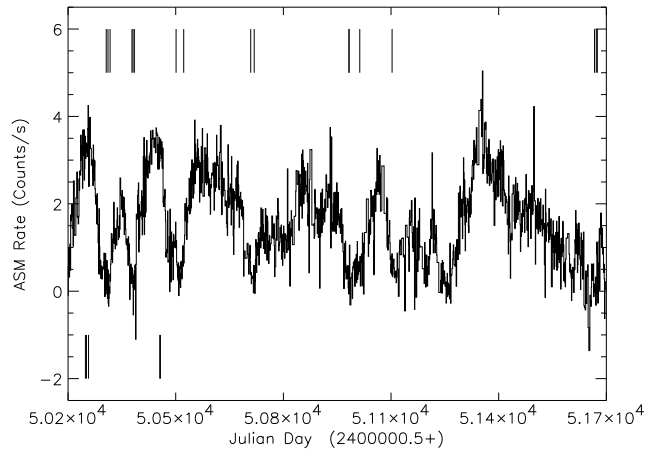,width=\hsize,angle=90}}
\caption{ Long term variation of LMC X-3 as observed by the
RXTE All-Sky Monitor.  One-day averages are shown.
PCA observations discussed
in this paper are indicated with vertical lines.  Target of Opportunity
Observations occurred near the minimum of the most recent ASM
minimum.
}
\end{center}
\end{figure}

For each of
the three 2000 May observations, 
the spectra were well modeled with a pure power
law, with photon index $\Gamma$=1.6-1.7. The derived 2--10~keV flux,
on the other hand, varies by a factor $\sim$3, between 4.7 and
15.7$\times$10$^{36}$ erg~s$^{-1}$ (at 50~kpc).  Table 1 summarizes
the observations and spectral fitting results; the single-component
power law model is sufficient to describe the spectrum, without the
need for a second continuum component or a line feature. (Wilms et
al.\ 2000 and Nowak et al.\ 2000 formally include such a line in their
fits but quote only upper limits on its detection.  
We derive an upper limit of 90 eV (90\% confidence) for
a Gaussian line with a width of 0.1 keV centered at 6.4 keV, 
comparable to the upper limit of 60 eV quoted by Wilms
et al.\ 2000).

In Figure 2 we show the data and derived model for the three
observations from 2000 May. We also include for comparison a spectrum
of \source\ taken at the relatively high ASM rate of 3 counts/s on
1996 April 28.  The high/soft spectrum is well described by the
power-law plus black body model with $\Gamma$=4.95 and
a disk inner-edge temperature of $kT$=1.34 keV.  

During the 3.5-yr period
covered by previous RXTE monitoring campaigns, \source\ has displayed
seven episodes where the count rate decreased to levels comparable
to those seen in 2000 May (Figure 1). To compare with the current
observations, we extracted PCA pointings from the RXTE archive
obtained near low ASM count rates.  
Many of the data sets are either too
short for good statistics, or are not centered in the minimum of
the low/hard state. In addition, the observations span all four gain
epochs of the PCA instrument.  We restricted ourselves to observations
containing at least 600s of good data, from PCA gain epochs 3 and 4
(1996 April 15 and following) where the calibration is best
understood.  We further limited ourselves to those minima where the
1-day ASM count rate was $<$0.5 counts/sec for more than 10 days.

The selected observations are included in Table~1, and their times
marked on Figure~1.  The data from 1998 May 29 are presented in Wilms
et al. (2000) as evidence for the low/hard state in \source, reanalyzed here
with the latest backgrounds and response matrices; the remainder of
the observations are previously unpublished.  For each observation,
the spectrum is well described by a featureless power law,
with photon index $\approxlt$1.8.  We conclude that these low
count-rate episodes have all the spectral characteristics of the
low/hard state.  
The spectrum of the low/hard state is characterized by a pure power law
with nearly constant photon index of $\sim$1.7$\pm$0.2, over a broad
range of flux corresponding to $L_x$=(2--16)$\times$10$^{36}$
erg~s$^{-1}$ at 50~kpc.  

\section{Timing Results}

The combined 3-20 keV power spectrum for the three 2000 May RXTE
observations is shown in Figure 3 (crosses).  A QPO peak is evident in
the data, centered at $\sim$0.5~Hz.

We experimented with two models for the band-limited noise: a single
power law, and a broken power law (the most commonly used
model for Galactic BHCs in their low/hard state).  The QPO was
modeled with a Lorentzian. The broken power law fit yielded a break
frequency of 0.15$^{+0.14}_{-0.03}$ Hz, and power law indices of
0.0$^{+0.2}_{-0.3}$ ($\nu<\nu_{break}$) and 0.77$\pm$0.07
($\nu>\nu_{break}$). The strength of the broken power law was
18.3$\pm$1.3\% rms in the 0.01--1 Hz range, and 40$\pm$4\% rms in the
0.01--100 Hz range. We measured a
central QPO frequency of 0.46$\pm0.02$ Hz, a FWHM of
0.14$^{+0.06}_{-0.04}$ Hz, and a strength of 14.0$^{+2.1}_{-1.9}$\%
rms. The $\chi^2_{r}$ for this model is 1.1 (for 210 d.o.f.) and the
fit is shown as a solid line in Figure 3.  The significance of adding
the QPO component to the broken power-law model was assessed using the
standard F-test.  For this case, the F-statistic $F_{s}$
is 3.92, with probability
$P(F>F_{s})$=0.009.

The single power law fit yielded an index of 0.58$\pm$0.04 and
strengths of 13.9$\pm$1.2\% rms (0.01--1 Hz) and 39$\pm$4\% rms
(0.01--100 Hz) for the noise. With this as the underlying model, we
determined a QPO frequency of 0.46$\pm$0.02 Hz, a FWHM of
0.23$^{+0.07}_{-0.06}$ Hz, and an rms amplitude of
18.4$^{+2.2}_{-1.8}$\%. The $\chi^2_{r}$ for this model is 
1.15 (for 212 d.o.f.).
For this case, adding the QPO results in an
 F-statistic of 16.8, with $P(F>F_{s})$
=8$\times$10$^{-10}$.  

\begin{figure}[htb]
\figurenum{2}
\begin{center}
{\psfig{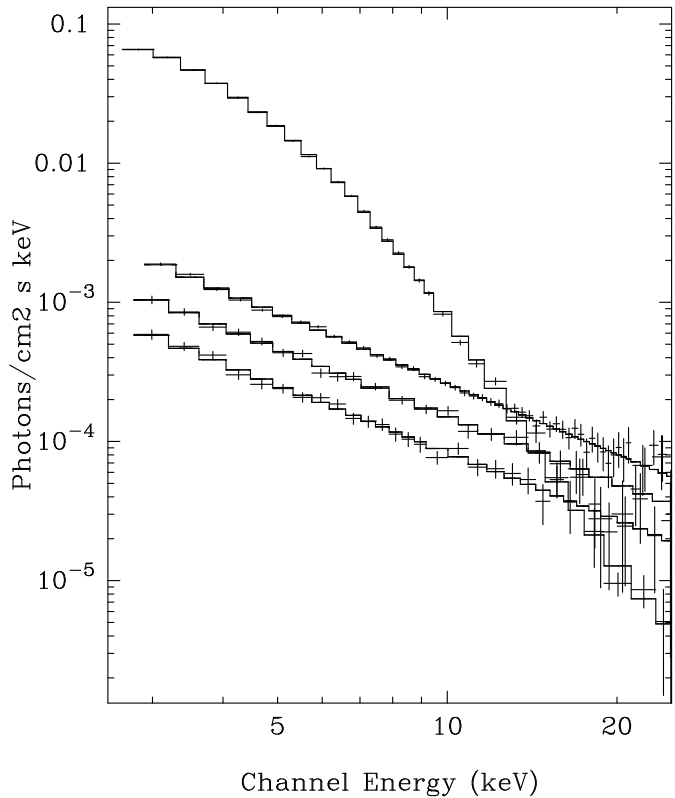}}
\caption{The PCA spectra from our three observations in 2000 May, together
with the pure power law fits.  The flux varies from 
4.7 to 15.7$\times$10$^{36}$ erg~s$^{-1}$ 
while the spectra are fit with a
constant photon index $\Gamma$=1.69$\pm$0.01.
A typical high/soft state spectrum from 1996 April 28 is shown for comparison.
Here, the disk + black body model is an acceptable fit, with 
photon index of 4.95 and a disk inner-edge
temperature of 1.34 keV.  The flux for this observation is 2.95$\times$10$
^{38}$ erg~s$^{-1}$. 
}
\end{center}
\end{figure}

\begin{figure}[htb]
\figurenum{3}
\begin{center}
\hspace{0.5cm}
{\psfig{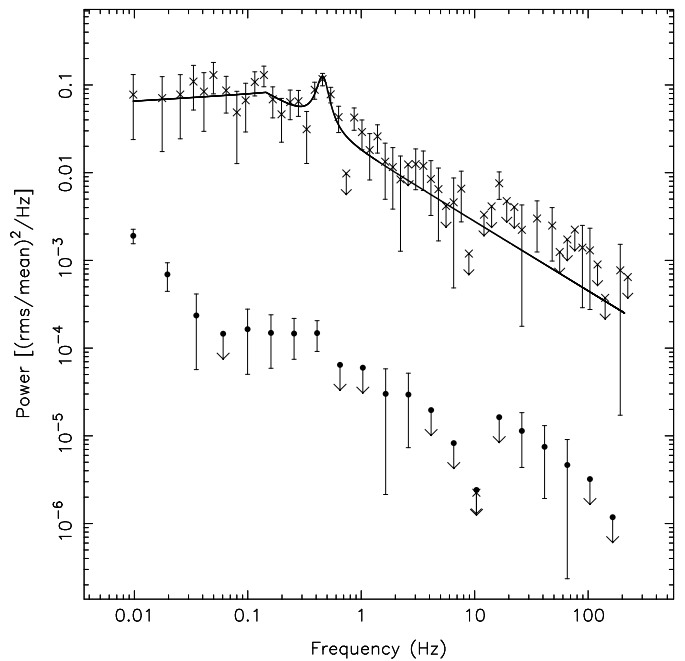}}
\caption{The combined 3-20 keV power spectra of LMC X-3 for the three
2000 May (crosses) and the November 30/December 2 1996 (bullets)
RXTE/PCA observations. For plotting purposes additional rebinning
was applied to the power spectra.  The solid line represents the best
fit with a Lorentzian and a broken power law (see Section 4 for
parameters). The arrows are 1 sigma upper limits to the power density.
The lower power spectrum is mostly due to background fluctiations and
should be regarded as an upper limit to the intrinsic source variability
in the high/soft state.}
\end{center}
\end{figure}

To study the energy dependence of the band-limited noise and the QPO,
we analyzed the power spectra in the 3-10 keV and 10-20 keV energy
bands.  We adopted the single power law for simplicity, fixed the
index, QPO frequency, and QPO FWHM to values obtained in the 3--20 keV
range, and allowed the other parameters to float.  In the 3--10 keV
band the rms amplitudes were 13.4$\pm$1.1\% (0.01--1 Hz), 38$\pm$3\%
(0.01--100 Hz), and 15.2$\pm$1.2\% (QPO); in the 10--20 keV band they
were 17$\pm$4\% (0.01--1 Hz), 50$\pm$12\% (0.01--100 Hz), and $<$24\%
(QPO).

We also compared the strength of the noise in the three 2000 May
observations with that of several archival RXTE/PCA observations 
during troughs and peaks in the ASM lightcurve (see Figure
1). The noise had strengths of 5--10\% rms (0.01--1
Hz) and 10--40\% rms (0.01--100 Hz) during the troughs,
somewhat lower than during the 2000 May observations but still
consistent with a low/hard state.  In the peak observations values
were found of $\sim$1\% rms (0.01--1 Hz) and $\sim$1.5\% rms
(0.01--100 Hz).  We compared the power spectra of the peak
observations with power spectra calculated from $\sim$25 ks randomly
chosen background observations, and concluded that the power spectra
of the peaks are consistent with those and should therefore be regarded
as upper limits (see also Nowak et al. 2000). The upper limits are
consistent with the source being in a high/soft state.  An example of
a power spectrum during high ASM count rate is included in Figure 3
(bullets). It is the combined power spectrum of the 1996 November 30
and December 2 observations (Nowak et al., 2000).

\section{Discussion}

The Galactic BHCs share many 
characteristics: (1) A mass function that implies a compact object
mass in excess of 3 $M_{\sun}$; (2) At least two distinct
emission states; 
(3) Prominent
time variability in the low/hard state, in the form of band-limited
noise with rms amplitudes of 30-50\%, in contrast to the weak
($<$10\%) variability seen in the higher states; and (4) QPO activity
in the range 0.01--10~Hz. (Useful summaries and references to results
from individual sources can be found in reviews by e.g.\ Tanaka \&
Lewin, 1995; van der Klis 1995; Wijnands \& van der Klis 1999.)
Until the current work, the extragalactic binary LMC~X--3
fulfilled only the first two of these criteria.  Here, we have
unambiguously determined the presence of the latter two
characteristics in LMC~X--3, which thus now joins the Galactic sources
in showing the full range of canonical BHC behavior.

At low inferred accretion rates, observed QPO frequencies in GBHCs fall
in the lower range 0.2--3~Hz (Wijnands \& van der Klis 1999, and references
therein). A total of eight GBHC exhibit both strong band-limited noise
{\sl and} low-frequency QPO peaks at these moderate to low accretion
rates; at frequencies $>$1~Hz, this noise component can be modeled as
a power law with index $\sim$1, with a break frequency at
$\sim$0.02--0.4~Hz, below which the spectral index flattens to
$\sim$0. For these sources, Wijnands \& van der Klis 1999 find a monotonic
relation between the QPO centroid frequency and the break frequency of
the band-limited noise.  Our measured values for the break frequency
(0.15~Hz) and the QPO frequency (0.46~Hz) in LMC~X--3 obey this
relation, suggesting that the basic (and unknown) physical mechanism
that underlies the fast aperiodic variability in the Galactic sources
extends also to LMC~X--3.

A total of ten GBHC sources with previously published QPOs and
corresponding spectral states were investigated by DiMatteo \& Psaltis
(1999), who conclude that the inner radii of the accretion disks
around the black holes do not change significantly from one state to
the next.  This is contrary to the qualitative predictions of the ADAF
models (Esin et al., 1998, and references therein), wherein the inner
radius retreats quite dramatically from soft-to-hard state
transitions.  DiMatteo \& Psaltis find that the GBHCs occupy a fairly
narrow, confined region when plotted in the photon index $\Gamma$
versus QPO frequency plane.  Our measured values for the photon index
(1.69) and QPO frequency (0.45~Hz) in \source\ fall in this region as
well.

Below a critical source luminosity L$<$ 5--10\%
Eddington, GBHCs have spectra described by a pure power law (Nowak,
1995).  Our observations, combined with the archival results presented
here, show that \source\ follows this trend, with the three low/hard
states presented here corresponding to a luminosity of $\sim$2\% or
less of $L_{Edd}$.

For all low/hard state properties measured here--photon index, QPO
frequency, band-limited noise, and luminosity-- \source\ falls squarely
in the range measured for the GBHCs.  This is significant, for it
implies that the dominant mechanism responsible for the low/hard
state and state transitions in BHCs is robust against
variations in system parameters such as compact object mass,
inclination, and initial chemical composition.

\section{Acknowledgments}

This paper utilizes quicklook results made publically available by the
ASM/RXTE Team, including members at MIT and NASA/GSFC, and also data
obtained through the High Energy Astrophysics Science Archive Research
Center Online Service, provided by the NASA/Goddard Space Flight
Center.  We thank Mariano M\'endez for useful discussions and his
help. We also acknowledge helpful conversations with Mike Nowak and
J\"orn Wilms, who shared their previous RXTE results with us prior to
publication.

\end{document}